\documentclass[prc,aps,twocolumn]{revtex4}
\usepackage{graphicx}% Include figure files
\usepackage{dcolumn}% Align table columns on decimal point
\usepackage{bm}% bold math
\usepackage{slashed}
\usepackage{ulem}
\usepackage{hyperref}
\usepackage{color}
\usepackage{url}
\begin{document}

\title{Searching for three-nucleon short-range correlations}
\author{Misak~M.~Sargsian$^{1}$, Donal~B.~Day$^{2}$,  Leonid~L.~Frankfurt$^{3}$, and Mark~I.~Strikman$^{4}$ }
\affiliation{
$^1$ Department of Physics, Florida International University, Miami, FL 33199, USA\\
$^2$ Department of Physics, University of Virginia, Charlottesville, VA 22904, USA\\
$^3$Sackler School of Exact Sciences, Tel Aviv University, Tel Aviv, 69978, Israel \\
$^4$ Department of Physics, Pennsylvania State University, University Park, PA 16802}

\date{\today}

\begin{abstract}
Three nucleon short range correlations~(SRCs) are one of the most elusive structures in nuclei. Their  observation and the subsequent study of 
their  internal makeup  will  have a significant impact on our understanding of the dynamics of super-dense nuclear matter 
which exists at the cores of neutron stars. We discuss the kinematic conditions and observables that are most favorable for probing 3N-SRCs in 
inclusive electro-nuclear processes and make a  prediction for a quadratic dependence of the probabilities of finding a nucleon in  
2N- and 3N- SRCs. We demonstrate that this prediction is consistent with the limited  high energy experimental data available, 
suggesting  that  we have observed, for the first time, 3N-SRCs in electro-nuclear processes. 
Our analysis enables us to extract $a_3(A,Z)$, the probability of finding 3N-SRCs in nuclei relative to the A=3 system. 
\\
\end{abstract}
\maketitle

\medskip
\medskip

\section{Introduction:}
Three nucleon short-range correlations~(3N-SRCs), in which three nucleons come close together, are unique arrangements in strong interaction 
physics. 3N~SRC's   have a single nucleon with very large momentum ($\gtrsim 700$~MeV/c)  balanced by two nucleons of comparable momenta.
Unlike two-nucleon short-range correlations~(2N-SRCs), 3N-SRCs have never been  probed directly  through experiment.
 As the consequence of the factorization  of   short-distance effects from low momentum collective phenomena~\cite{FS81,srcrev},
2N- and 3N- SRCs dominate the  high momentum component of nuclear wave function which is almost universal   
up to a scale factor~(see e.g.\cite{FS81,Atti:2015eda}).

The dynamics of  three-nucleon  short-range  configurations reside at the borderline of our knowledge of nuclear forces  making their exploration a testing ground for  ``beyond the standard nuclear physics" phenomena such as irreducible three-nucleon forces, inelastic transitions in 3N systems as well as the transition from hadronic to quark degrees of freedom.   Their strength is expected to grow faster with the local nuclear density than 
the strength  of 2N-SRCs~\cite{FS81,srcrev}.  
As a result, their contribution  will be essential  for an understanding of the dynamics of super-dense nuclear matter (see e.g. Ref.~\cite{PanHei:2000}).

Until recently a straightforward experimental  probe of 2N- and 3N-SRCs  was impossible due to  the  requirements  of high-momentum transfer nuclear  reactions being measured in very specific  kinematics  in which  the expected cross sections are very small~(see Ref.\cite{FS81} and references therein).  With the advent  of the high energy (6~GeV) and high intensity  continuous electron accelerator at Jefferson Lab~(JLab)  in the late 1990's,  an  unprecedented  exploration of   nuclear structure became possible, opening a new window to multi-nucleon SRCs.

\begin{figure}[th]
\vspace{-0.3cm}
\includegraphics[height=2.4cm,width=5.4cm]{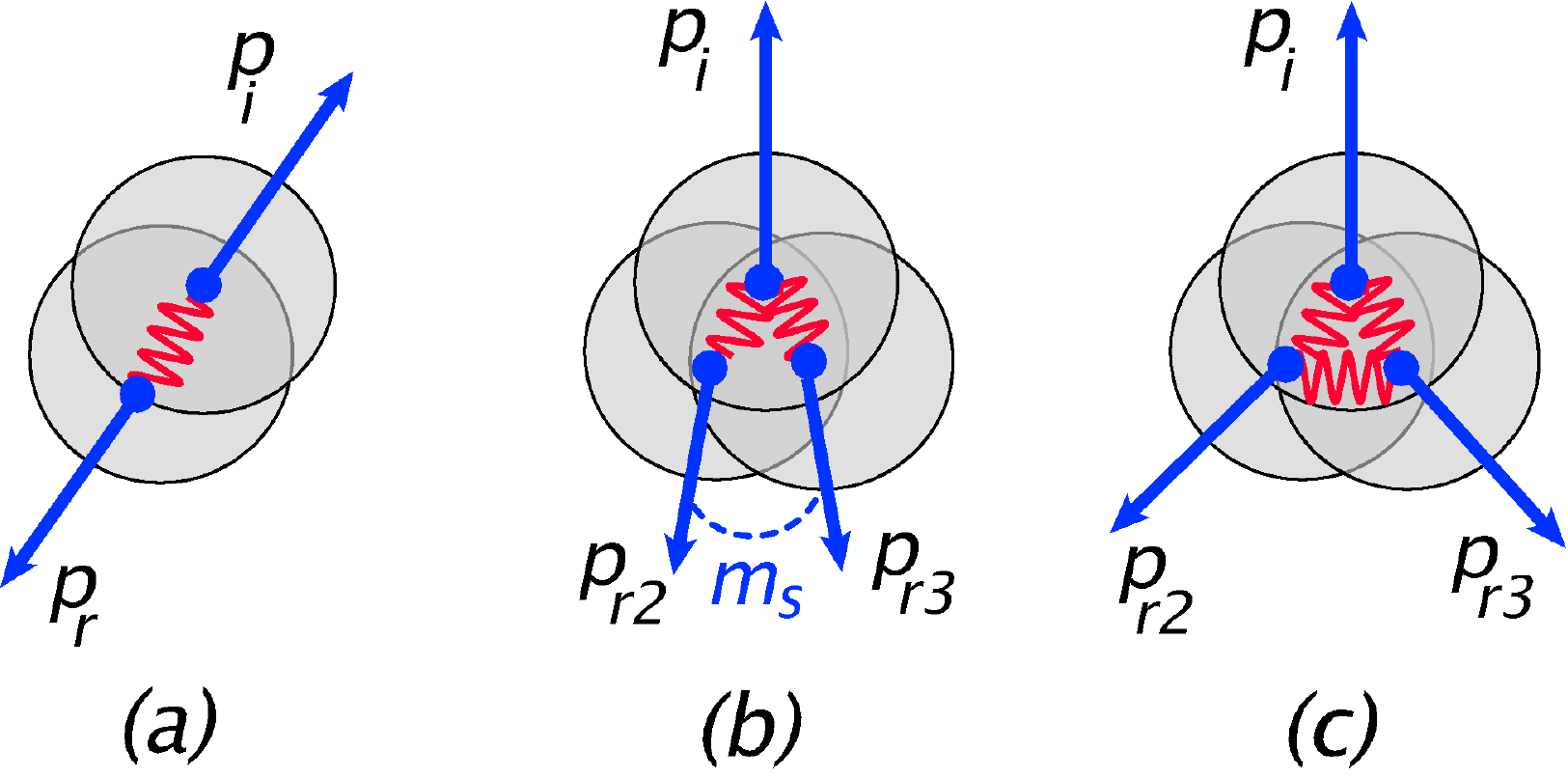}
\vspace{-0.3cm}
\caption{(a) Geometry of 2N-SRCs, ${\bf p_r}\approx-{\bf p_i}$. 
Two configurations of 3N-SRCs: (b) Configuration in which recoil nucleon momenta $\bf {p_{r2}}, \bf {p_{r3}} \sim {\bf - {p_i}/2}$, 
(c) configuration in which $p_{r2}\sim p_{r3}\sim p_i$. Here $m_s$ is the invariant mass of the recoiling 2N system.}
\label{src}
\vspace{-0.2cm}
\end{figure}

\section{ Two Nucleon Short Range Correlations~(2N-SRCs)}
 The first dedicated study of  2N-SRCs  in  inclusive, $A(e,e^\prime)X$, high momentum transfer reactions  revealed a plateau  in the ratios  of  per nucleon cross sections of heavy nuclei to  the deuteron \cite{FSDS} measured at Stanford Linear Accelerator Center~(SLAC) with momentum transfer, $Q^2\gtrsim 2$~GeV$^2$ and Bjorken variable $x> 1.5$. Here  $x = {Q^2\over 2m_N q_0}$ with $m_N$  the nucleon mass and $q_0$ the transferred energy to the nucleus, and for a nucleus $A$, $0< x < A$. The observed plateau, largely insensitive to $Q^2$ and $x$,  sets the parameter $a_{2}(A,Z)$\cite{FS88} which is the probability of finding 2N-SRCs in the ground state of the nucleus A relative to the deuteron.   These plateaus were confirmed  in 
 inclusive cross section ratios of nuclei A to  $^3$He\cite{Kim1,Kim2},  at similar kinematics with the magnitude of  plateaus taken to be related to the relative probability, ${a_{2}(A,Z)\over a_{2}(^3He)}$. Qualitatively and quantitatively the latter results were in agreement with Ref.\cite{FSDS}.  These, together with  more recent and dedicated measurements of the nuclear to the deuteron inclusive cross section ratios\cite{Fomin2011}, provided compelling evidence for the sizable ($\sim 20\%$)  high momentum component of the ground state nuclear wave function  for medium to heavy nuclei originating from 2N-SRCs.

While inclusive processes provided the first evidence for 2N-SRCs and an estimate of their probabilities, $a_2(A,Z)$,  the details of correlation dynamics have been obtained 
mainly from semi-inclusive experiments in which one  or both  nucleons  from 2N-SRCs  were detected.
The first $A(p,ppn)X$ experiments at high momentum transfer  were performed at Brookhaven National Laboratory\cite{eip1,eip2}.
The theoretical analysis of these experiments  gave the striking result that the probability of finding proton-neutron combinations  in 2N-SRCs exceeds by almost a factor of 20 the probabilities for proton-proton and neutron-neutron SRCs\cite{isosrc}.  
This  result was subsequently   confirmed in semi-inclusive electroproduction reactions at JLab\cite{eip3,eip4} and both are understood as arising from the dominance of the tensor component in the NN interaction at distances  $|r_1 -r_2|  \lesssim 1$~fm \cite{eheppn2,Schiavilla:2006xx}. This reinforced the conclusion that  the nucleons have been isolated in SRCs with separations much smaller than average inter-nucleon distances. The dominance of the $pn$ component in 2N-SRCs suggested a new  prediction for  momentum sharing between high momentum protons and neutrons in asymmetric nuclei\cite{newprops} where the minority component (say protons in neutron rich nuclei) will dominate  the high momentum component of the nuclear wave function. This prediction was confirmed indirectly in $A(e,e'p)X$ and $A(e,e'pp)X$  experiments\cite{twofermi} and directly in $A(e,e'p)X$ and $A(e,e'n)X$ processes in which proton and neutron constituents of 2N-SRCs have been probed independently\cite{pndirect,Duer:2018sxh}.   The momentum sharing effects also arise from variational Monte-Carlo calculations for light asymmetric nuclei\cite{Wiringa} as well as in  model calculations of nuclear wave functions for medium to heavy nuclei\cite{Vanhalst:2014cqa}.

In addition to measuring the isospin content of 2N-SRCs, several experimental analyses\cite{eip2,eip3,Erez18} established a detailed ``geometrical" picture of 
2N-SRCs consisting of  two  overlapping nucleons having relative momentum between $250-650$~MeV/c with back-to-back angular correlations (Fig.{\ref{src}(a))  
and with moderate center of mass momentum,  $\lesssim 150$~MeV/c, for nuclei ranging from $^4$He to $^{208}$Pb\cite{Erez18}. 
Several recent reviews\cite{srcrev,srcprog,Atti:2015eda,arnps,Hen:2016kwk}  have documented extensively the recent progress  in the investigation of 2N-SRCs in a wide range of atomic nuclei.

\section{Three Nucleon Short Range Correlations~(3N-SRCs)} 
Despite an impressive progress achieved in studies of 2N-SRCs the confirmation of 3N-SRCs remains  arguable.  One signature of 3N-SRCs is the onset  of the {\it plateau}  in the ratio of inclusive cross sections of nuclei A and $^3$He in the kinematic region of $x>2$ similar to  the  plateau  observed for 2N-SRCs in the region of $1.5 < x < 2$  and discussed above. The first   observation of  a  plateau at $x>2$ was claimed in Ref.\cite{Kim2}. However it was not confirmed  by subsequent measurements\cite{Fomin2011,Ye:2017mvo}. The source of this  disagreement has been traced to the poor resolution at $x>2$ of the experiment of Ref.\cite{Kim2} which led to bin migration\cite{Higinbotham:2014xna} where events move  from smaller to higher $x$.  
Additionally, as it will be shown below, the  absence  of a plateau is related to the the modest invariant momentum transfer, $Q^2$ covered in Ref.~\cite{Kim2}.

\begin{figure}[tb]
\vspace{-0.7cm}
\includegraphics[height=7cm,width=9cm]{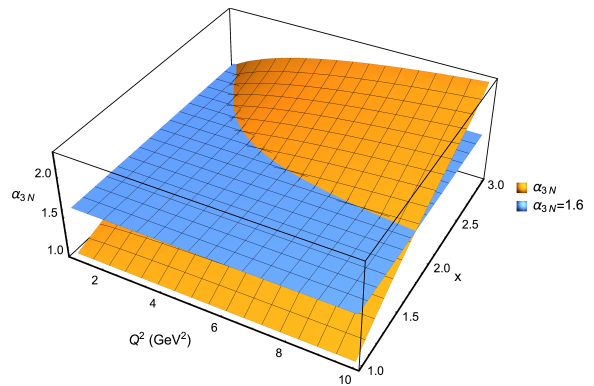}
\vspace{-1.0cm}
\caption{Kinematics of 3N-SRCs. The surface above the horizontal plane at $\alpha_{3N}=1.6$ defines the kinematics most optimal for 
identification of 3N-SRCs in inclusive processes. In this calculation we assumed a minimal mass for $m_S= 2m_N$ which corresponds to 
the maximal contribution to the nuclear spectral function with $k_\perp=0$ and $\beta = 1$ (see Eq.(\ref{alpha3N})).}
\vspace{-0.5cm}
\label{3Nkins}
\end{figure}

To quantify  the last statement we first need to identify the dominant structure of 3N-SRCs in the nuclear ground state.  The problem is that while for 2N-SRCs the correlation geometry is straightforward (two fast nucleons nearly balancing each other, Fig.\ref{src}(a)), in the case of 3N-SRCs the geometry 
of balancing three fast nucleons is not unique - ranging from configurations in which two almost parallel spectator nucleons with momenta, 
$\sim {\bf - {p_i\over 2}}$  balance the third nucleon with momentum ${\bf p_i}$, Fig.\ref{src}(b)),  to the configurations in  which all  three nucleons have momenta $p_i$ with relative angles $\approx 120^0$
 Fig.\ref{src}(c)). The analysis of  3N systems in Ref.\cite{eheppn2} demonstrated that configurations  in which two recoil nucleons have the smallest possible  mass, $m_S\sim 2m_N$, dominate the 3N-SRC nuclear spectral function at lower excitation energy.  
This allows us to conclude \cite{DFSS18} that in  inclusive scattering,  which integrates over the nuclear excitation energies, the dominant contribution to  3N-SRCs originates from  arrangements similar to Fig.\ref{src}(b)) with $m_S\gtrsim 2m_N$.

With the dominant mechanism of 3N-SRCs identified we are able to develop the kinematic requirements to expose 3N correlations in inclusive $eA$ scattering.  We use the fact that, due to relativistic nature of SRC configurations,  the most natural description   is achieved through the light-cone~(LC) nuclear spectral functions\cite{FS88,multisrc} in which the correlated nucleons are described by their nuclear  light-cone momentum fractions, $\alpha_i$ and transverse momentum $p_{i,\perp}$.   In inclusive scattering one probes the  spectral function integrated over the LC momenta of the correlated recoil nucleons, residual nuclear excitation energy  and the   transverse momentum  of the interacting nucleon. 
This corresponds to the LC density matrix of the nucleus $\rho_A(\alpha_N)$, 
where $\alpha_N$ is the LC momentum fraction of the nucleus carried by the interacting nucleon. It can be shown\cite{FSS15}  that $\rho_A(\alpha_N)/\alpha$ is analogous to the partonic distribution function in QCD, $f_i(x)$ where $x$  describes the LC momentum fraction of the nucleon carried by the interacting quark.
 
To  evaluate the LC momentum fraction, $\alpha_{N}$ (henceforth $\alpha_{3N}$)  describing the interacting nucleon in the 3N-SRC, 
we consider the kinematic condition of quasielastic scattering from  a 3N system:  $q + 3m_N = p_f + p_{S}$, where $q$, $p_f$ and $p_S$ are the four momenta of 
the virtual photon, final struck nucleon and recoil two-nucleon system respectively.  One defines the LC momentum fraction of  the interacting nucleon, $\alpha_{3N} = 3- \alpha_S$, 
where $\alpha_S \equiv 3{E_S - p_S^z\over E_{3N}-p_{3N}^z}$ is the light-cone fraction of the two spectator  nucleons in the center of mass of the $\gamma^* (3N)$ system with 
$z || q$. Using the boost invariance of 
the light-cone momentum fractions one arrives at the following lab-frame expression (see Ref.\cite{DFSS18} for details) :
\begin{eqnarray}
& & \alpha_{3N}   = 3 \ - \ {q_- + 3 m_N\over 2 m_N} \left[1 \ \  + \ \ {m_S^2 - m_N^2\over W_{3N}^2} \ \ +  \right. \ \ \  \nonumber \\  
  & & \left.    
\sqrt{\left(1 - {(m_S + m_N)^2\over W_{3N}^2}\right)\left(1 - {(m_S - m_N)^2\over W_{3N}^2}\right)}\right],
\label{alpha3N}
\end{eqnarray}
where  $W^2_{3N} = (q + 3m_N)^2 = Q^2{3-x\over x} + 9 m_N^2$ and  $q_- = q_0 -  q$ with $q_0$ and $q$ being energy and momentum 
transfer in the lab with $z|| {\bf q}$. 
The invariant mass of the spectator 2N system, $m_S^2 = 4{m_N^2 + k_\perp^2\over \beta (2-\beta)}$ where
$\bf {k_\perp}$ is the transverse component of the relative momentum of the 2N system with respect  to $\bf {p_S}$ 
and $\beta$ is the light-front momentum fraction of $p_S$ carried by  the spectator nucleon ($0\le \beta\le 2$). 
Inclusive reactions integrate over the nuclear spectral function and $k_\perp$ and $m_s$ are not determined experimentally.

 The expression for $\alpha_{3N}$, Eq.~(\ref{alpha3N}),  makes it possible to identify the kinematical conditions most appropriate for the isolation of  3N-SRCs in 
inclusive $A(e,e^\prime)X$ reactions.  This is done by identifying the minimal value of $\alpha_{3N}$ above which one expects 
the  contribution of  3N-SRCs to dominate. First, the threshold can be established from our experience of studying 2N-SRCs.
 In  this  case we know that  2N-SRCs in inclusive processes  dominate at $\alpha_N\ge 1.3$  which corresponds to an  internal longitudinal  momenta of 
$\sim 300-350$~MeV/c. 
Hence for 3N-SRCs one needs at least $p_{min} \gtrsim 700$~MeV/c, corresponding to $\alpha_{3N} \gtrsim  1.6$,
which will allow two  high momentum  spectator nucleons to belong to a  3N-SRCs. 
This minimal value for $\alpha_{3N}$ is validated by  the studies  of the fast backward nucleon production in pA scattering within the few-nucleon correlation 
model~\cite{FS88} which indicate that the transition from 2N- to 3N- SRCs  occurs at $\alpha \sim 1. 6 - 1.7$.
 
As $\alpha_{3N}$ increases  above $1.6$  the contribution of 2N-SRCs is suppressed relative to 3N-SRCs. 
 This is because as the  LC momentum fraction grows, the relative momentum in the 2N system grows much faster than the same quantity in the 3N system. 
 Thus, in the further discussions we will set 
$\alpha_{3N}=1.6$ as the threshold value,  above which one expects the 3N-SRCs to dominate in inclusive scattering.  
This minimal value for $\alpha_{3N}$ allows us to identify the kinematic parameters most promising for probing 3N-SRCs as illustrated in Fig.~\ref{3Nkins}. The figure shows the relevant kinematics corresponding to the $\alpha_{3N}$ surface being above the $\alpha_{3N}=1.6$ plane.  This identifies the $Q^2$ and $x$ domains favorable  for  probing 3N-SRCs.  
In particular, one observes that starting around $Q^2\gtrsim 2.5-3$~GeV$^2$ one gains enough  kinematical range  in the  $x>2$ domain that one 
expects to observe  3N-SRCs. 

Another advantage of considering 3N-SRCs in terms of $\alpha_{3N}$,  is that at sufficiently large $Q^2$ the LC momentum 
distribution function $\rho_A(\alpha_{3N})$ is not altered 
by final state  hadronic interactions~(FSIs). The important feature in the high energy limit is that FSIs redistribute  the $p_\perp$ strength in the 
nuclear spectral function leaving $\rho_A(\alpha_{3N})$  practically unchanged\cite{ms01,ggraphs,BS15}.  In this limit 
the   distortion  of  $\alpha_{3N}$ due to FSI can be evaluated as\cite{ms01}: 
\begin{equation} 
\delta \alpha \approx {x^2\over Q^2}{2m_N E_R\over (1 + {4m_N^2x^2\over Q^2})},
\label{dalpha}
\end{equation}
where $E_R$ is the kinetic energy of the recoil two nucleon system.
The estimates  made in Ref.\cite{DFSS18} indicate that for $Q^2 \sim 3$~GeV$^2$ FSI may alter  $\alpha_{3N}$  by not more than 
$8\%$ which is too small to shift the mean field nucleon, $\alpha_N \approx 1$, to the 3N-SRC domain at $\alpha_{3N}\ge 1.6$. 

\section{Signatures of 3N-SRCs} 
The cross section in inclusive electron scattering at high $Q^2$ is factorized in the form\cite{FS88}: 
\begin{equation}
\sigma_{eA} \approx \sum\limits_N \sigma_{eN}\rho^{N}_A(\alpha_N),
\label{eA}
\end{equation}
where $\sigma_{eN}$ is the  elastic  electron-bound nucleon scattering cross section and $\rho^N_A(\alpha_N)$ is the
light-front density matrix of the  nucleus at a given LC momentum fraction, $\alpha_N$ of the probed nucleon.
This is analogous to the QCD factorization in  inclusive deep-inelastic scattering off the nucleon, in which the cross section is a product of 
a hard   electron-parton scattering cross section and  partonic distribution function.

The local property of  SRCs  suggests that $\rho_A(\alpha_N)$  in the correlation region to be proportional to the light-front density matrix of the two- and three-nucleon systems\cite{FS88,FSDS}. This expectation leads to the prediction of the plateau for the ratios of inclusive cross sections in the SRC region that has been confirmed for  2N-SRCs. Similar to 2N-SRCs for the 3N-SRC  one predicts  a plateau for the experimental cross section ratios such as:
\begin{equation}
R_{3}(A,Z) = {3\sigma_{A}(x,Q^2)\over A \sigma_{^3He}(x,Q^2)}\left|_{\alpha_{3N}>\alpha^0_{3N}}\right.,
\label{R3}
\end{equation}
 where  $\alpha^0_{3N}$  is the threshold value for the $\alpha_{3N}$ above  which one expects onset of 3N-SRCs (taken as $\sim 1.6$ as described 
 above). 
To  quantify  the strength of  3N-SRCs we introduce  a parameter $a_3(A,Z)$\cite{DFSS18}:
\begin{equation}
a_3(A,Z) = {3\over A} {\sigma_{eA}\over (\sigma_{e^3He} + \sigma_{e^3H})/2},
\label{a3}
\end{equation}
representing an intrinsic nuclear property related to the probability of finding 3N-SRCs in the nuclear ground state.  
If a plateau is observed in the 3N-SRC region of $\alpha_{3N}$  then the ratio $R_{3}(A,Z)$ in Eq.(\ref{R3}) can be used to extract 
$a_3(A,Z)$ as follows\cite{DFSS18}:
\begin{equation}
a_3(A,Z) =  R_3(A,Z) {(2\sigma_{ep} + \sigma_{en})/3\over (\sigma_{ep}+\sigma_{en})/2}.
\label{a3_R3}
\end{equation}

The status of the experimental observation of the scaling in the ratio of Eq.(\ref{R3}) is as follows:
The E02-109 experiment\cite{Arrington:2002ex}  provided  a high accuracy 
ratios, in the 2N-SRC region, at large momentum transfer for a wide range of nuclei\cite{Fomin2011}. 
This experiment 
covered some part of the 3N-SRC kinematic region with lesser quality of data (see also Refs.\cite{Arrington:2002ex,Fomin:2010ei,FominAIP,FominPHD}),
providing an indication  of a plateau in the cross section ratios beginning at $x>2$
once  $Q^2$ is sufficiently high.

 In Ref.\cite{DFSS18} it was pointed out that the above-mentioned data \cite{Arrington:2002ex,Fomin2011,Fomin:2010ei} suffered from a collapse of 
 the $^3He$ cross section between $x = 2.68$ and $x = 2.85$ due to difficulties with the subtraction of the Aluminum target walls. 
 This issue arose from the relatively small diameter of the target cell (4~cm) combined with  the fact that $\sigma^{\rm{Al}} \gg \sigma^{^3\rm{He}}$ 
 at large $x$ as $\sigma^{^3\rm{He}}$ must go to 0 at its kinematic limit, $x=3$.  
The cross section ratio  in Ref.~\cite{Fomin2011} was made possible by the 
following: First the inverted ratio $^3$He/$^4$He was formed and then rebinned - combining three bins into one for $x \geq 1.15$. 
Subsequently the bins in the inverted ratio that had error bars falling below zero were moved along a truncated gaussian, such that the lower edge of the error bar was at zero.  
The ratio was then inverted to give the ratio for $^4$He/$^3$He shown in Figure 3 of Ref.~\cite{Fomin2011} and as the  triangles in 
Fig.~\ref{ratio4to3} below. The use of a truncated gaussian gave rise to the asymmetric error bars seen 
in the ratios.

As an alternative to 
the somewhat unconventional procedure above, we have 
used the following approach to substitute the $^3$He data of Refs.\cite{Arrington:2002ex,Fomin2011,Fomin:2010ei} in 3N-SRC region:
We have replaced the problematic data  between $x = 2.68$ and $x = 2.85$  $(1.6 \leq \alpha_{3N} \leq 1.8)$, point by point, 
by employing a y-scaling function $F(y)$\cite{Day:1987az,Sick:1980ey,Day:1990mf} fit to  the SLAC data \cite{Day:1979bx,Rock:1981aa}  
measured at a comparable  $Q^2$.  
 A simple,  two parameter fit  $F(y) = a \exp({-bx})$,  limited to the range 
$1.6 (y =-0.7)\le \alpha_{3N} \le 1.8 (y = -1.1)$ provides a good description of the the SLAC data\cite{DFSS18}.
We preserved the  absolute error of the E02019 data set \cite{Arrington:2002ex,Fomin2011,Fomin:2010ei} 
rather than the smaller errors from the fit. The fit parameters  are $a =0.296$ and $b = 8.241$.

\begin{figure}[th]
%\vspace{-1.2cm}
\includegraphics[width=0.5\textwidth]{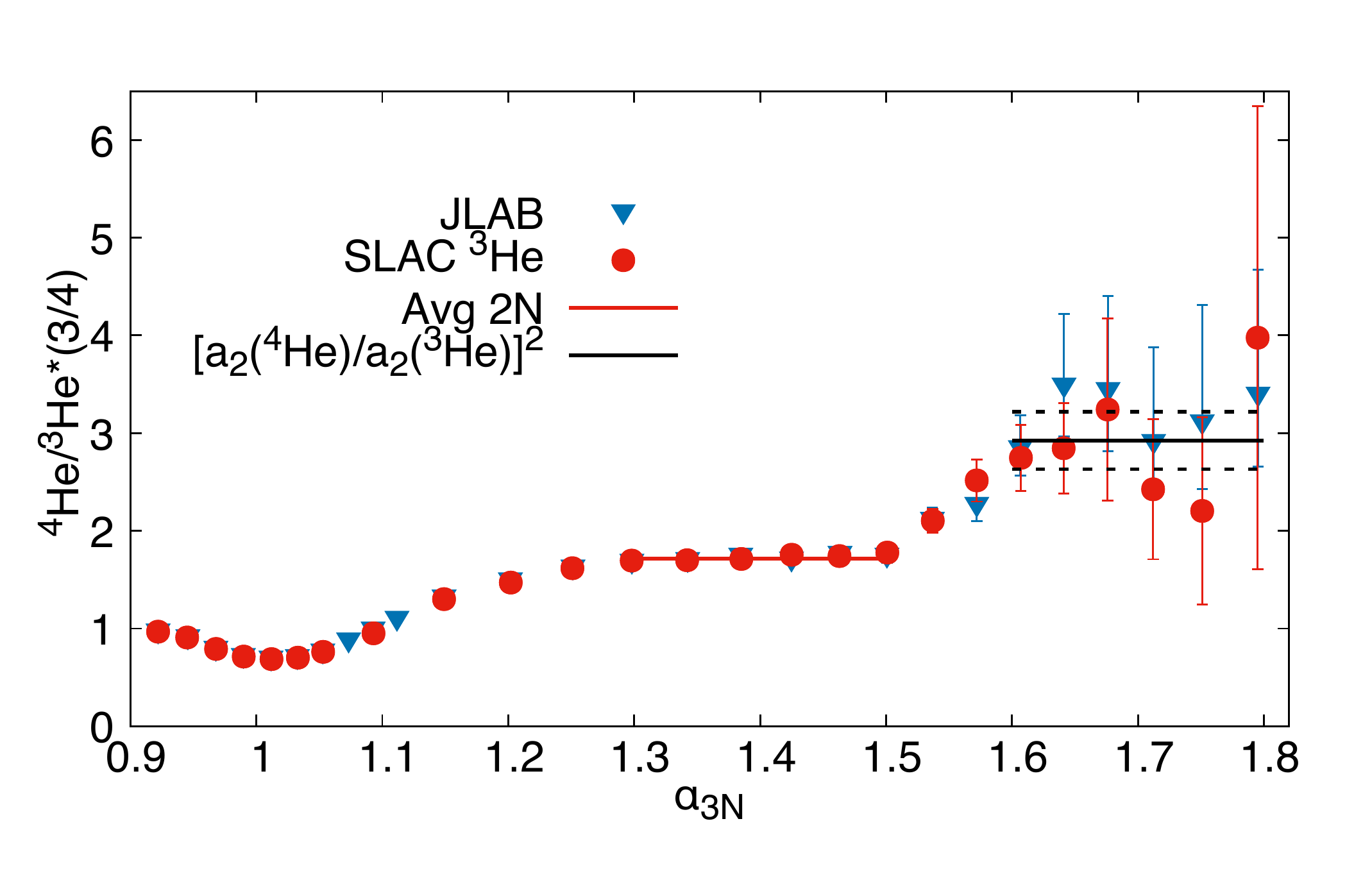}
\vspace{-0.5cm}
\caption{The $\alpha_{3N}$ dependence of the inclusive cross section ratios for $^4$He to  $^3$He, triangles - JLAB data~\cite{Fomin2011,Fomin:2010ei}, 
circles - ratios  when using a parameterization of SLAC $^3$He cross sections~\cite{Day:1979bx,Rock:1981aa}. 
The horizontal line at $1.3\le \alpha_{3N}<1.5$ identifies the magnitude of the 2N-SRC plateau. 
The line for $\alpha_{3N}>1.6$ is Eq.(\ref{R3_to_R2sq}) with a $10\%$ error introduced to account for the systematic uncertainty in $a_2(A,Z)$ parameters across
all measurements. The data correspond to $Q^2\approx 2.5$~GeV$^2$ at $x=1, \alpha_{3N} =1$.}
\label{ratio4to3}
\end{figure}

Note  that the above approach was  first used in Ref.~\cite{FSDS},
which provided the first evidence of 2N-SRCs through cross section ratios in inclusive scattering.
The 2N-SRC results  obtained have been confirmed by subsequent precision studies\cite{Kim1,Kim2,Fomin2011}  
in which the ratios were measured in single experiment. 
 
 It is also worth mentioning  that  in the case of 2N-SRC the adopted approach was more complicated than the one we employed in the current work.
In Ref.~\cite{FSDS} the data were combined to form the cross section ratios of nuclei ($^3$He, $^4$He, C, Al, Fe and Au) to the deuteron, covering  
a range in Q$^2$ from 0.9 to 3.2 (GeV/c)$^2$.
In the current analysis of 3N-SRCs,  we worked at a single value of Q$^2 \approx$ 2.7 (GeV/c)$^2$ and, incidentally, the $^3$He data used in 1993 is the same set we employ here. The resulting ratios are displayed as red circles in Fig.~\ref{ratio4to3}.  

Fig.~\ref{ratio4to3} presents the results  for the cross section ratios obtained within the two above described approaches.
While both give similar results we consider 
the replacement of the data points between $x = 2.68$ and $x = 2.85$ $(1.6 \leq \alpha_{3N} \leq 1.8)$ as a best alternative 
to the procedure adopted in Ref~\cite{Fomin2011} in part because it allows a consistent treatment of the ratios for all $A $.

In Fig.~\ref{ratio4to3} the plateau 
due to 2N-SRCs is clearly visible   for $1.3 \le \alpha_{3N} \le 1.5$.  
In this region $\alpha_{3N}\approx \alpha_{2N}$\cite{DFSS18}, where $\alpha_{2N}$ is the LC momentum fraction of the nucleon in 
the  2N-SRC. Because of this, we refer to   the magnitude of this plateau as:
\begin{equation} 
R_2(A,Z) =  {3\sigma_{A}(x,Q^2)\over A \sigma_{^3He}(x,Q^2)} \left|_{1.3\le \alpha_{3N}\le 1.5}\right. = \frac{a_2(A)}{a_2(^3He)}.\label{eq:R2}
\end{equation}
The horizontal line in the region of $1.3 \le \alpha_{3N} \le 1.5$ is  given by the right hand side of  Eq.~(\ref{eq:R2}), in which  the values of 
 $a_2(^3He)$ and  $a_2(A)$  are taken from the last column of Table II in Ref.~\cite{Arrington:2012ax}, an average of the SLAC, JLAB Hall C and JLAB Hall B results. The magnitude of the  horizontal solid line in the region of  $1.6 \le \alpha_{3N} \le 1.8$, is  the prediction 
 of $R_{3N}(A,Z)\approx R_{2N}^2(A,Z)$ 
which will be explained in the next section.  We assigned a $10\%$ error to this prediction (dashed lines) related to the uncertainty of 
$a_2(A,Z)$ magnitudes across different measurements.

As Fig.\ref{ratio4to3}  shows the data at $\alpha_{3N}>1.6$  are consistent with the prediction of the onset of the new plateau in the 3N-SRC region and
that  its magnitude}  is proportional to $R_{2N}^2$.  
 
With a set of $^3$He data 
obtained  in the above discussed approach
we are able to estimate the ratios for other nuclei, including, $^9$Be, $^{12}$C, $^{64}$Cu, and $^{197}$Au, albeit with larger uncertainties\cite{DFSS18}.   

The large experimental uncertainties in evaluation of the ratios for $^4He$ (Fig.\ref{ratio4to3}) and for heavier nuclei\cite{DFSS18} do not allow us to 
claim unambiguously the onset of the plateau at $\alpha_{3N}\ge 1.6$.  However one can evaluate the validity of such a plateau by comparing 
one- and two- parameter  fits  to the experimental ratios in the $\alpha_{3N}\ge 1.6$ region.
The  one-parameter fit in the 3N-region gives the values ($R_3^{exp}$) of the plateaus as seen in Figure~\ref{ar3ar2}(a) along with our prediction of Eq.~(\ref{R3_to_R2sq}). $R_3^{exp}$ is also listed in Table~\ref{a2R3a3}.
A two-parameter linear fit, returns errors on the parameters nearly as large as the parameters themselves and a correlation matrix 
indicating that the second parameter is  redundant,  providing no additional information.

\section{3N-SRCs and the  $\bf pn$  Dominance:} 
\label{pndominance}
In Fig.\ref{src}(b) the  3N-SRC is  produced in the sequence  of  two short-range $NN$ interactions in which the nucleon with the 
largest momentum interacts with the  external probe\cite{multisrc,DFSS18}. 
The presence of  short-range $NN$ interactions in  3N-SRC configurations  tells  us that  the   recently observed $pn$-SRC 
dominance\cite{isosrc,eip4,eip3} is critical to our understanding of 3N-SRCs.

For 3N-SRCs one expects that only $pnp$ or $npn$ configurations to contribute,   with the $pn$ short-range 
interaction  playing role of  a ``catalyst" in forming a fast interacting nucleon with momentum, $p_i$~(Fig.\ref{src}(b) ).
For example, in the case of $pnp$ configuration, the  neutron will play the role of intermediary in furnishing a large momentum transfer 
to one of the protons with two successive  short range $pn$ interactions.  
Quantitatively such a scenario is reflected in  the nuclear light-front density matrix in the 3N-SRC domain,
$\rho^N_{A(3N)}(\alpha_N)$, being expressed through the convolution of two $pn$-SRC density matrixes, $\rho^N_{A(pn)}(\alpha,p_\perp)$ 
as follows:
\vspace{-0.3cm}
 \begin{eqnarray}
 \rho^N_{A(3N)}(\alpha_N,p_\perp)  \approx    \sum\limits_{i,j}\int F(\alpha^\prime_i,p_{i\perp},\alpha^\prime_j,p_{j\perp}) \times \ \ \ \ \ \ \ \ \ \ \ &&
\nonumber \\  
 \rho^N_{A(pn)}\left(\alpha^\prime_i,p^\prime_{i\perp}\right) \rho^N_{A(pn)}\left(\alpha^\prime_j,p^\prime_{j\perp}\right)  d\alpha_id^2p_{j\perp}
d\alpha_id^2p_{j\perp}, &&
\label{rho3}
\end{eqnarray}
where $(\alpha^\prime_{i/j}, p^\prime_{i/j\perp})$,  are the LC momentum fractions and transverse momenta of spectator nucleons in the center of mass of  the $pn$ SRCs.   
According to the $pn$ dominance\cite{newprops}: 
\begin{equation}
 \rho^N_{A(pn)}(\alpha,p_\perp) \approx {a_2(A,Z)\over 2X_N} \rho_d(\alpha,p_\perp),
\label{prop2}
\end{equation}
 where $X_N = Z/A$ or $(A-Z)/A$ is the relative fraction of the proton or neutron in the nucleus and $\rho_d(\alpha,p_\perp)$ is the light-front density function of the deuteron at $\alpha\ge 1.3$. The factor $F(\alpha^\prime_i,p_{i\perp},\alpha^\prime_j,p_{j\perp})$ is a smooth function of LC momenta and accounts for the phase factors of nucleons in the intermediate state between the sequential $pn$ interactions with $0< \alpha_{i/j}^\prime<2$.

\begin{figure}[th]
 %\vspace{-3.2cm}
\includegraphics[width=0.45\textwidth]{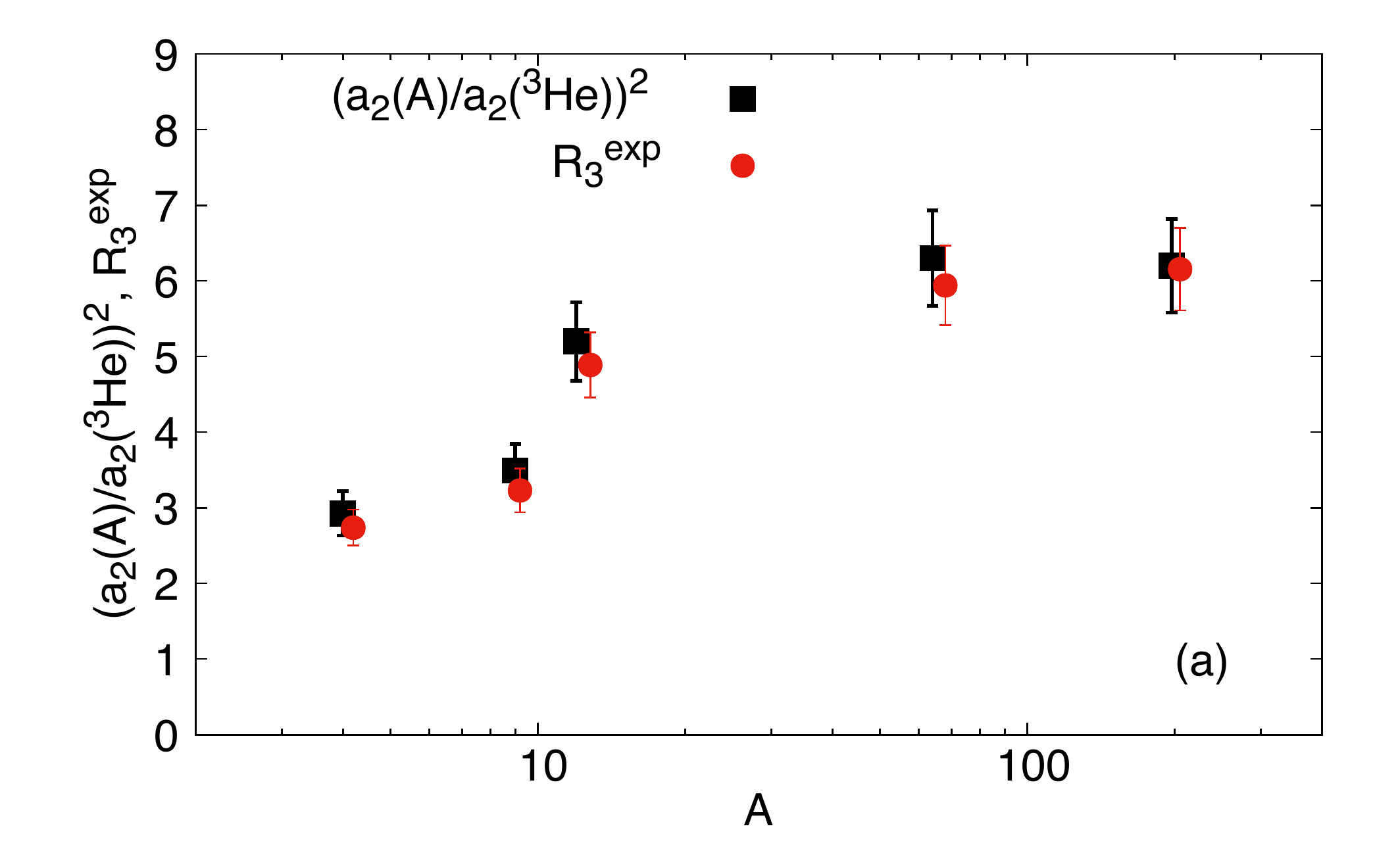}
\includegraphics[width=0.45\textwidth]{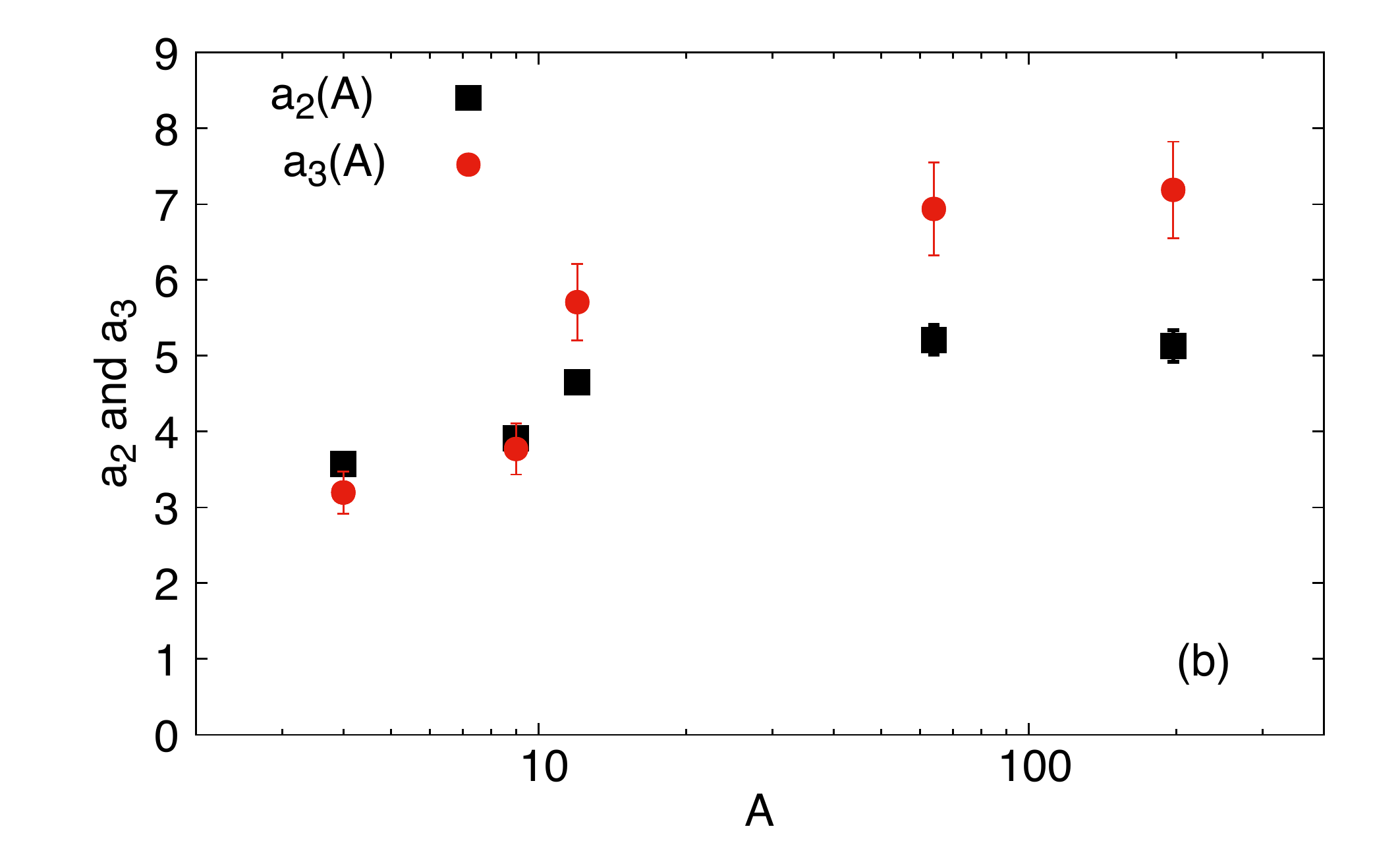}
%\vspace{-0.5cm}
\caption{(a) The $A$ dependence of the experimental evaluation of $R_3$ compared with the 
prediction of Eq.\ref{R3_to_R2sq}. (b) The $A$ dependence of 
$a_3(A,Z)$ parameter compared to $a_2(A,Z)$ of Ref.\cite{Fomin2011}. }
\label{ar3ar2}
\vspace{-0.5cm}
\end{figure}

It follows, from Eq.(\ref{rho3}) and the expression of  $\rho^N_{A(pn)}(\alpha,p_\perp)$ in Eq.(\ref{prop2}),  that  the strength of  3N-SRCs is $\propto a_2^2(A,Z)$. This is  evident by calculating 
$R_3$ in Eq.(\ref{R3}) using the relation~(\ref{eA}) and the conjecture of Eq.(\ref{rho3}), which  leads to\cite{DFSS18}: 
\begin{equation} R_3(A,Z) =   {9\over 8}{(\sigma_{ep}+\sigma_{en})/2\over (2\sigma_{ep} + \sigma_{en})/3} 
R^2_2(A,Z)  \approx \left( {a_2(A,Z)\over a_2(^3He)}\right)^2, 
\label{R3_to_R2sq} 
\end{equation} 
where $\sigma_{ep} \approx 3\sigma_{en}$  in  the considered $Q^2\sim$ 3~GeV$^2$ range. As Fig.\ref{ratio4to3} shows the prediction of $R_3 \approx R_2^2$ is in agreement with the experimental per nucleon cross section ratios of $^4$He to  $^3$He targets. There is a similar agreement  for other nuclei including  $^9$Be, $^{12}$C, $^{64}$Cu and $^{197}$Au\cite{DFSS18}.

\begin{table}[h]
\centering
\caption{Numerical values a$_2$\cite{Arrington:2012ax}, R$_2$ (Eq.~\ref{eq:R2}), R$_3^\textrm{exp}$ (the weighted average in the 3N region)  and  $a_3$ (Eq.~\ref{a3_R3})).}
\vspace{0.2cm}
\begin{tabular}{|c|c|c|c|c|}\hline
A   & a$_2$    & $R_2$       &  $R_3^{\textrm{exp}}$ &  a$_3$ \\ \hline
3   & 2.13 $\pm 0.04$ &  1  &   NA                     &    NA	   \\ \hline
4   & 3.57 $\pm 0.09$ & $1.68 \pm  0.03$  &  2.74 $\pm 0.24$  &  $3.20 \pm 0.28$  \\ \hline
9   & 3.91 $\pm 0.12$ & $1.84 \pm  0.04$  &  3.23 $\pm 0.29$  &  $3.77 \pm 0.34$  \\ \hline
12  & 4.65 $\pm 0.14$ & $2.18 \pm  0.04$   & 4.89 $\pm 0.43$   &  $5.71 \pm 0,50$   \\ \hline
64  & 5.21 $\pm 0.20$ & $2.45 \pm  0.04$   & 5.94 $\pm 0.52$   &  $6.94 \pm 0.77$   \\ \hline
197 & 5.13 $\pm 0.21$ & $2.41 \pm  0.05$  & 6.15 $\pm 0.55$   &  $7.18 \pm 0.64$   \\ \hline
\end{tabular}
\vspace{0.2cm}
\label{a2R3a3}
\end{table}

To test  the prediction of Eq.(\ref{R3_to_R2sq}) quantitatively 
we evaluated the weighted average of $R^{exp}_3(A,Z)$ for $\alpha_{3N}> 1.6$ and compared them with the magnitude of  
$({a_2(A,Z)\over a_2({3He})})^2$ in which $a_2(A,Z)$'s are taken from  Ref.~\cite{Arrington:2012ax}. The results  in which the $^3$He 
cross section was taken from the $F(y)$ fit to the SLAC data  are presented in  Fig.\ref{ar3ar2}(a) and in Table~\ref{a2R3a3}. 
They show   good agreement with the prediction of Eq.(\ref{R3_to_R2sq}) for the full  range of nuclei. 
We investigated the sensitivity of the weighted average of $R_3(A,Z)$ on the lower limit of $\alpha_{3N}$ (before rebinning) 
and found that the results shown in Fig.~\ref{ar3ar2}(a) remain unchanged within errors which 
grow with a larger $\alpha_{3N}>1.6$ cut.

The agreement presented in Fig.\ref{ar3ar2}(a) represents the strongest evidence yet for  the presence of 3N-SRCs.
If it  is truly due to the onset of  3N-SRCs then one can use the measured $R^{exp}_3$ ratios and Eq.(\ref{a3_R3}) to extract the $a_3(A,Z)$ 
parameters characterizing the 3N - SRC probabilities in the nuclear ground state. The estimates of $a_3(A,Z)$ and comparisons with 
$a_2(A,Z)$ are  given in Fig.\ref{ar3ar2}(b) (see also Table~\ref{a2R3a3}). 
These comparisons show a faster rise for $a_3(A,Z)$  with $A$, consistent with the expectation 
of the  increased sensitivity of 3N-SRCs to  the local  nuclear density\cite{srcrev}.  If this result is verified in the future with better quality 
data and a wider range of nuclei  then the evaluation of the parameter $a_3(A,Z)$ as a function of nuclear density and 
proton/neutron asymmetry together with $a_2(A,Z)$  can provide an important theoretical input for the  exploration of the dynamics of super dense nuclear matter (see e.g. \cite{Ding:2016oxp}).

\section{Summary} Based on the theoretical analysis of a three-nucleon system we  have concluded that the dominating mechanism 
of 3N-SRCs in inclusive processes corresponds to the situation in which the recoil mass of the 2N spectator system 
is close to a minimum.   From that basis we derived a kinematic condition for the onset of 3N-SRCs in inclusive eA scattering  
which should result in the  observation of a plateau in the ratio of cross sections of heavy to light nuclei,  such as, $\frac{3}{A}\frac{\sigma^A}{\sigma^{3He}}$.  
The best quality data, available for large enough $Q^2$ (Fig.\ref{ratio4to3}), 
indicate a  possible onset of such a plateau at $\alpha_{3N}> 1.6$. This first signature of 3N-SRCs is reinforced by the 
good agreement with the prediction of the quadratic ($R_3\approx R^2_2$) dependence between the cross section ratios in the 3N-SRCs domain, 
$R_3$, and the same ratio  measured in the 2N-SRC region, $R_2$. This agreement has allowed us, for the first time, 
to extract the parameter $a_3(A,Z)$ characterizing the strength of 3N-SRCs in the ground state wave function of the nucleus. 
Further measurements at larger $Q^2$ are necessary to confirm the observation made in this  analysis. 
Precision data at large $Q^2$  in the 3N-SRC region  can be secured  in the  forthcoming  12~GeV experiment at Jefferson Lab, 
E12-06-105\cite{Arrington:2006xx}.

\begin{acknowledgments} 
This work is supported by the US Department of Energy grants: 
DE-FG02-96ER40950 (DBD), DE-FG02-01ER41172 (MSS) and DE-FG02-93ER40771 (MIS). 
\end{acknowledgments}
%We would like to thank Oscar Rondon for useful guidance when treating the data.

%%%%%%%%%%%%%%%%%%%%%%%%%%%%%%%%%%%%%%%%%%%%%%%%%%%%%%%%%%%%%%%%%%%%%%%%%

\end{document}